# Spectral redshift of the thermal near field scattered by a probe

Sheila Edalatpour[*,†], Vahid Hatamipour[**] and Mathieu Francoeur[**,‡]

[*]Department of Mechanical Engineering, University of Maine, Orono, ME 04469, USA

[**]Radiative Energy Transfer Lab, Department of Mechanical Engineering, University of Utah,

Salt Lake City, UT 84112, USA

**ABSTRACT**

The physics underlying spectral redshift of thermally generated surface phonon-polaritons (SPhPs) observed in near-field thermal spectroscopy is investigated. Numerically exact fluctuational electrodynamics simulations of the thermal near field emitted by a silicon carbide surface scattered in the far zone by an intrinsic silicon probe show that SPhP resonance redshift is a physical phenomenon. A maximum SPhP redshift of 19 cm$^{-1}$ is predicted for a 200-nm-diameter hemispherical probing tip and a vacuum gap of 10 nm. Resonance redshift is mediated by electromagnetic gap modes excited in the vacuum gap separating the probe and the surface when the probing tip is much larger than the gap size. The impact of gap modes on the scattered field can be mitigated with a probing tip size approximately equal to or smaller than the vacuum gap. However, sharp probing tips induce important spectral broadening of the scattered field. It is also demonstrated that a dipole approximation with multiple reflections cannot be used for explaining the physics and predicting the amount of redshift in near-field thermal spectroscopy. This work shows that the scattered field in the far zone is a combination of the thermal near field emitted by the surface, and electromagnetic interactions between the probe and the surface. Spectroscopic analysis of near-field thermal emission thus requires a numerically exact fluctuational electrodynamics framework for modeling probe-surface interactions.

---

[†] Corresponding author (S. Edalatpour). Tel.: + 1 207 581 2375
Email address: sheila.edalatpour@maine.edu
[‡] Corresponding author (M. Francoeur). Tel.: + 1 801 581 5721
Email address: mfrancoeur@mech.utah.edu

Several recent experiments [e.g., 1-6] have reported enhancement of near-field radiative heat transfer beyond the blackbody limit between two objects separated by a sub-wavelength vacuum gap, as measured [7] and later theoretically explained using fluctuational electrodynamics [8] more than four decades ago. In addition, it is well known that thermal excitation of resonant electromagnetic surface modes characterized by high degrees of spatial and temporal coherence can drastically modify near-field thermal emission spectra [9,10]. For instance, quasi-monochromatic near-field thermal emission and radiative transfer can be achieved with materials such as silicon carbide (SiC) and doped silicon (Si) respectively supporting surface phonon-polaritons (SPhPs) and surface plasmon-polaritons in the infrared [11,12]. The ability to probe near-field thermal spectra is of critical importance because several potential applications, such as thermophotovoltaics [13-15] and thermal rectification [16-18], capitalize on spectrally selective near-field radiative exchange. Yet, to date, only a handful of articles have reported spectroscopic measurements of near-field thermal emission by planar heat sources [19-23]. In near-field thermal spectroscopy, the evanescent component of the thermal field is scattered in the far zone using a probing tip brought within a sub-wavelength distance from a heat source. The far-zone scattered field is guided to a Fourier-transform infrared spectrometer, where its spectral distribution is extracted. The collected signal is interpreted as a measurement of the spectral near-field energy density of the heat source. However, all near-field thermal spectroscopy experiments performed with heat sources supporting SPhPs in the infrared (SiC, silicon dioxide, hexagonal boron nitride) have reported resonance redshifts of varying magnitude, from ~ 5 cm$^{-1}$ to ~ 65 cm$^{-1}$, using probing tips made of intrinsic Si [19,21], tungsten [20] and platinum-iridium [22]. The mechanism causing SPhP resonance redshift, experimental or physical, is still unclear. Several simplified models, namely the point dipole model with [20,24] and without retardation effects [19], the finite dipole model [21,25], and the effective medium theory [21] have been used in an attempt to reproduce resonance redshift arising in near-field thermal spectroscopy. These methods rely on adjustable parameters (e.g., dipole size) to fit theoretical predictions with experimental data, and therefore do not provide physical insight into resonance redshift.

In this paper, a critical knowledge gap is addressed by analyzing the physics of the near-field thermal interactions between a probe and a surface via the thermal discrete dipole approximation (T-DDA) [26,27], which is a numerically exact method based on fluctuational electrodynamics. By considering a planar heat source made of SiC and a non-resonant probe made of intrinsic Si, it is shown that SPhP resonance redshift observed on the scattered field in the far zone is not an experimental artifact but is a physical phenomenon mediated by electromagnetic gap modes generated in the vacuum gap separating the probe and the surface. This work shows unambiguously that the far-zone scattered field is not a sole signature of near-field thermal emission by the surface, but is also greatly affected by the probe-surface electromagnetic interactions.

The problem under consideration is shown in Fig. 1(a), where a probe of intrinsic Si couples evanescent modes emitted by a planar heat source of SiC into propagating modes detectable in the far zone. This choice is motivated by the experiments reported in Ref. [21] where spectral redshifts varying from 5 cm$^{-1}$ to 50 cm$^{-1}$ with respect to SPhP resonance of a SiC-vacuum interface were observed on the field scattered by different intrinsic Si probes. In all simulations, the vacuum gap $d$ is fixed at 10 nm, and the surface ($T_s$) and probe ($T_p$) temperatures are respectively 573 K and 0 K. A probe temperature of 0 K is selected since intrinsic Si is nearly transparent in the infrared such that its thermal emission at room temperature is negligible compared to emission by the SiC surface. The dielectric function of intrinsic Si is taken from Ref. [28], while the dielectric function of SiC is modeled using a Lorentz oscillator: $\varepsilon_s(\omega) = \varepsilon_\infty[(\omega^2 - \omega_{LO}^2 + i\Gamma\omega)/(\omega^2 - \omega_{TO}^2 + i\Gamma\omega)]$, where $\varepsilon_\infty = 6.7$, $\omega_{LO} = 969$ cm$^{-1}$, $\omega_{TO} = 793$ cm$^{-1}$, and $\Gamma = 4.76$ cm$^{-1}$ [28].

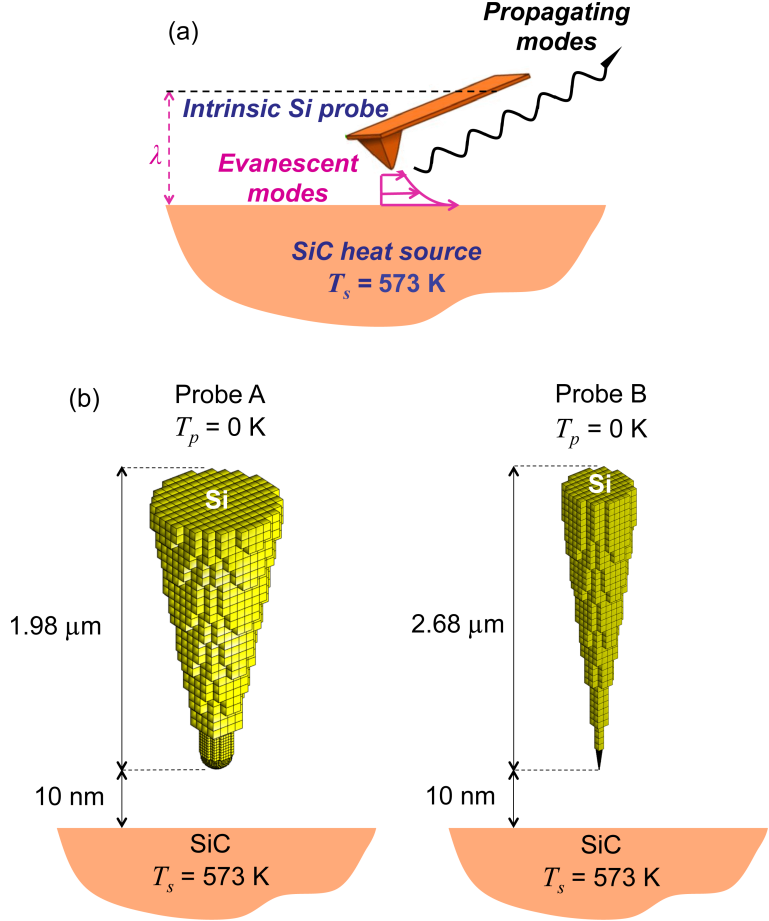

FIG. 1. (a) Schematic representation of the problem under consideration. The intrinsic Si probe couples evanescent modes emitted by the SiC surface into propagating modes detectable in the far zone. (b) Discretized probes used in T-DDA simulations. Probe A is 1.98-μm-long, has an opening angle of 19°, and a 200-nm-diameter hemispherical tip. Probe B is 2.68-μm-long, has an opening angle of 12°, and a sharp conical tip having a size of 7.2 nm.

Near-field thermal interactions between the probe and the surface are calculated in a numerically exact manner via the T-DDA. In this framework, the probe is discretized into $N$ cubical sub-volumes much smaller than the material and vacuum wavelengths, and the vacuum gap (necessary only in a small portion of the probe facing the surface), such that they can be conceptualized as electric point dipoles [26]. The surface is modeled as an infinite plane and its interactions with the sub-volumes constituting the probe are treated analytically via Sommerfeld's theory of electric dipole radiation above an infinite plane [29,30]. Two probe geometries are considered in the simulations (see Fig. 1(b)). Probe A is 1.98-μm-long and is characterized by a 200-nm-diameter hemispherical tip, while probe B is 2.68-μm-long and is characterized by a needle-like, sharp conical tip with a size of 7.2 nm. The number and size of the sub-volumes used for discretizing the probes are reported in Section 1 (Table S1) of the Supplemental Material [31]. The sub-volumes in the probe are illuminated by a fluctuating electric field due to thermal emission by SiC. The resulting induced dipole moment $\mathbf{p}_i$ in a specific sub-volume $i$ depends not only on the fluctuating field, but also on the direct interaction with all other sub-volumes and on the indirect interaction with the sub-volumes after reflection at the vacuum-SiC interface. The induced dipole moments are calculated via the following system of equations [27]:

$$\left(\overline{\overline{\mathbf{A}}} + \overline{\overline{\mathbf{R}}}\right) \cdot \overline{\mathbf{P}} = \overline{\mathbf{E}}^{sur} \tag{1}$$

where $\overline{\mathbf{P}}$ is the $3N$ stochastic column vector containing the unknown induced dipole moments, $\overline{\mathbf{E}}^{sur}$ is the $3N$ stochastic column vector containing the fluctuating field due to thermal emission by the surface, $\overline{\overline{\mathbf{A}}}$ is the $3N$ by $3N$ deterministic interaction matrix calculated from the free-space dyadic Green's function (direct interaction between sub-volumes), and $\overline{\overline{\mathbf{R}}}$ is the $3N$ by $3N$ deterministic reflection-interaction matrix calculated from reflection dyadic Green's function (indirect interaction between sub-volumes). The scattered electric field (magnitude squared) in the far zone is obtained from the induced dipole moments as follows:

$$\left\langle \left| \mathbf{E}^{\text{FF}}(\mathbf{r}) \right|^2 \right\rangle = \frac{k_0^4}{16\pi^2 r^2} \text{tr}\left( \overline{\overline{\mathbf{F}}} \cdot \left\langle \overline{\mathbf{P}} \otimes \overline{\mathbf{P}} \right\rangle \cdot \overline{\overline{\mathbf{F}}}^{\dagger} \right) \tag{2}$$

where $k_0$ is the magnitude of the wavevector in vacuum, $r$ is the magnitude of the position vector $\mathbf{r}$, $\otimes$ is the outer product, and $\dagger$ denotes the Hermitian operator. The ensemble average of the induced dipole moment autocorrelation function is determined from Eq. (1):

$$\left\langle \overline{\mathbf{P}} \otimes \overline{\mathbf{P}} \right\rangle = \left( \overline{\overline{\mathbf{A}}} + \overline{\overline{\mathbf{R}}} \right)^{-1} \cdot \left\langle \overline{\mathbf{E}}^{sur} \otimes \overline{\mathbf{E}}^{sur} \right\rangle \cdot \left[ \left( \overline{\overline{\mathbf{A}}} + \overline{\overline{\mathbf{R}}} \right)^{-1} \right]^{\dagger} \tag{3}$$

An explicit expression for the ensemble average of the spatial correlation function of the fluctuating field, $\left\langle \overline{\mathbf{E}}^{sur} \otimes \overline{\mathbf{E}}^{sur} \right\rangle$, has been provided in Ref. [27]. In Eq. (2), $\overline{\overline{\mathbf{F}}}$ is the 3 by $3N$ deterministic matrix accounting for direct emission by the induced dipoles in the far zone (direct component) as well as radiation in the far zone after reflection of dipole emission at the vacuum-SiC interface (indirect component). The matrix $\overline{\overline{\mathbf{F}}}$, obtained using the far-zone approximation of the free space and reflection dyadic Green's functions [32], consists of $N$ 3 by 3 sub-matrices $\mathbf{F}_i$ given by:

$$\mathbf{F}_i = a_i^{\text{TM}}\left( \mathbf{e}_1 \otimes \mathbf{e}_1 \right) + a_i^{\text{TE}}\left( \mathbf{e}_2 \otimes \mathbf{e}_2 \right), \ i = 1,2,\ldots,N \tag{4}$$

In Eq. (4), $\mathbf{e}_1$ and $\mathbf{e}_2$ are unit vectors oriented along the transverse magnetic (TM) and transverse electric (TE) polarizations, respectively, and the coefficient $a_i^{\gamma}$ ($\gamma$ = TM or TE) is defined as:

$$a_i^{\gamma} = \exp\left( -i\mathbf{k}_{sca} \cdot \mathbf{r}_i \right) + R^{\gamma} \exp\left( -i\mathbf{k}_{sca} \cdot \mathbf{r}_{1,i} \right) \tag{5}$$

where $\mathbf{r}_i$ is the position vector of sub-volume $i$, $\mathbf{r}_{1,i}$ is the position vector of the image of sub-volume $i$ with respect to the vacuum-SiC interface, $\mathbf{k}_{sca}$ is the wavevector in the far zone, and $R^\gamma$ is the Fresnel reflection coefficient for $\gamma$-polarized radiation incident from vacuum.

The electric field scattered in the far zone by probes A and B is shown in Fig. 2 at an angle $\theta$ of 45° with respect to the surface normal and at a radial distance $r$ of 1 m away from the surface. The near-field thermal energy density in vacuum at a distance $\Delta$ of 10 nm above the SiC surface, calculated via fluctuational electrodynamics [33] and plotted for comparison, is quasi-monochromatic at a frequency of 948 cm$^{-1}$. This frequency corresponds to SPhP resonance of a SiC-vacuum interface, $\omega_{res} \approx \sqrt{(\varepsilon_\infty \omega_{LO}^2 + \omega_{TO}^2)/(\varepsilon_\infty + 1)}$, expression derived by determining the poles in the TM-polarized Fresnel reflection coefficient of the interface in the electrostatic limit and by neglecting losses in the dielectric function of SiC [34]. The spectral distribution of the field scattered by probe A is drastically different from the energy density spectrum. Indeed, the scattered field exhibits a resonance at a frequency of 929 cm$^{-1}$, which represents a redshift of 19 cm$^{-1}$ with respect to SPhP resonance. A local maximum at a frequency of 940 cm$^{-1}$ is also visible on the scattered field spectrum. Numerically exact simulations of near-field thermal interactions between a probe and a surface clearly demonstrate that SPhP resonance redshift is not an experimental artifact. The physics underlying SPhP resonance redshift can be explained by analyzing gap modes, which are electromagnetic modes existing in the vacuum gap separating the probe and the surface [35-39]. The eigenfrequencies of gap modes, predicted by Mal'shukov for a lossless sphere-surface configuration under the assumption that that $D \gg d$, where $D$ is the sphere diameter, are given by [37]:

$$\frac{1}{\varepsilon'_p(\omega)} + \frac{1}{\varepsilon'_s(\omega)} = -(2m+1)\sqrt{\frac{d}{D}}, \quad m = 0,1,2,... \tag{6}$$

where $\varepsilon'_p$ and $\varepsilon'_s$ are the real parts of the dielectric function of the spherical probe and the surface, respectively, while $m$ is the order of the gap mode. In addition to the aforementioned conditions, Eq. (6) shows that generation of gap modes requires the real part of the dielectric function of the surface and/or the probe to be negative. As such, gap modes cannot be generated between a non-resonant surface and a non-resonant probe.

As probe A is characterized by a 200-nm-diameter hemispherical tip, Eq. (6) is used for predicting gap modes with $D = 200$ nm. The resulting gap mode eigenfrequencies are 915 cm$^{-1}$, 942 cm$^{-1}$ and 951 cm$^{-1}$ for the zeroth, first and second order, respectively. Only the zeroth and first order gap modes are visible on the scattered field spectrum, and the predictions made with Eq. (6) are in reasonable agreement with the T-DDA simulations. Discrepancies are explained by the fact that while the probe of intrinsic Si is quasi-lossless in the infrared, SiC exhibits non-negligible losses ($\varepsilon'' \sim 0.2$ in the 915-930 cm$^{-1}$ spectral band). Conversely, the scattered field spectrum of probe B characterized by a sharp tip with size approximately equal to the vacuum gap thickness $d$ is not greatly impacted by gap modes. As such, a minor redshift of 4 cm$^{-1}$ is observed on the field scattered by probe B. It is also worth mentioning that the field scattered by probe B experiences an important spectral broadening, and is not dominated by SPhP resonance. A very small portion of the volume of probe B is in the extreme near field where SPhPs are dominant. Since the scattered field in the far zone is proportional to the volume squared of the illuminated portion of the probe, the scattered field is dominated by the contribution from the large portion of the probe volume that is located at micrometer distances away from the surface. Spectral broadening with respect to the near-field energy density observed with probe A is also mediated by the same physics. Note that the fields scattered by both probes display a local maximum at 980 cm[-1] due to a local maximum in the reflection by the surface at angle of 45°.

Results of Fig. 2 and Eq. (6) suggest that resonance redshift is a strong function of the probe geometry and probing tip size to vacuum gap thickness ratio. Probes used in near-field thermal spectroscopy have a complex shape and it is difficult to precisely quantify their dimensions. This explains the results reported by O'Callahan et al. [21] where SPhP resonance redshift of the scattered field of a SiC surface ranging from 5 cm[-1] to 50 cm[-1] was observed when the experiment was repeated using different intrinsic Si probes. SPhP resonance redshift obtained here falls within the range measured in Ref. [21]. In addition, it is interesting to note that the tip of probe A has approximately the same size as the tungsten probing tip used in the experiments conducted in Ref. [20]. Tungsten is characterized by a large negative real part of the dielectric function ($\sim -10^3$) in the infrared [40]. Assuming that probe A is made of tungsten, the zeroth order gap mode is predicted at a frequency of approximately 903 cm[-1], which results in a more severe redshift of 45 cm[-1]. This is in line with the redshift of 35 cm[-1] reported in Ref. [20] for a SiC surface. Therefore, metals having dielectric function with large negative real part in the infrared should be avoided if resonance redshift is to be minimized.

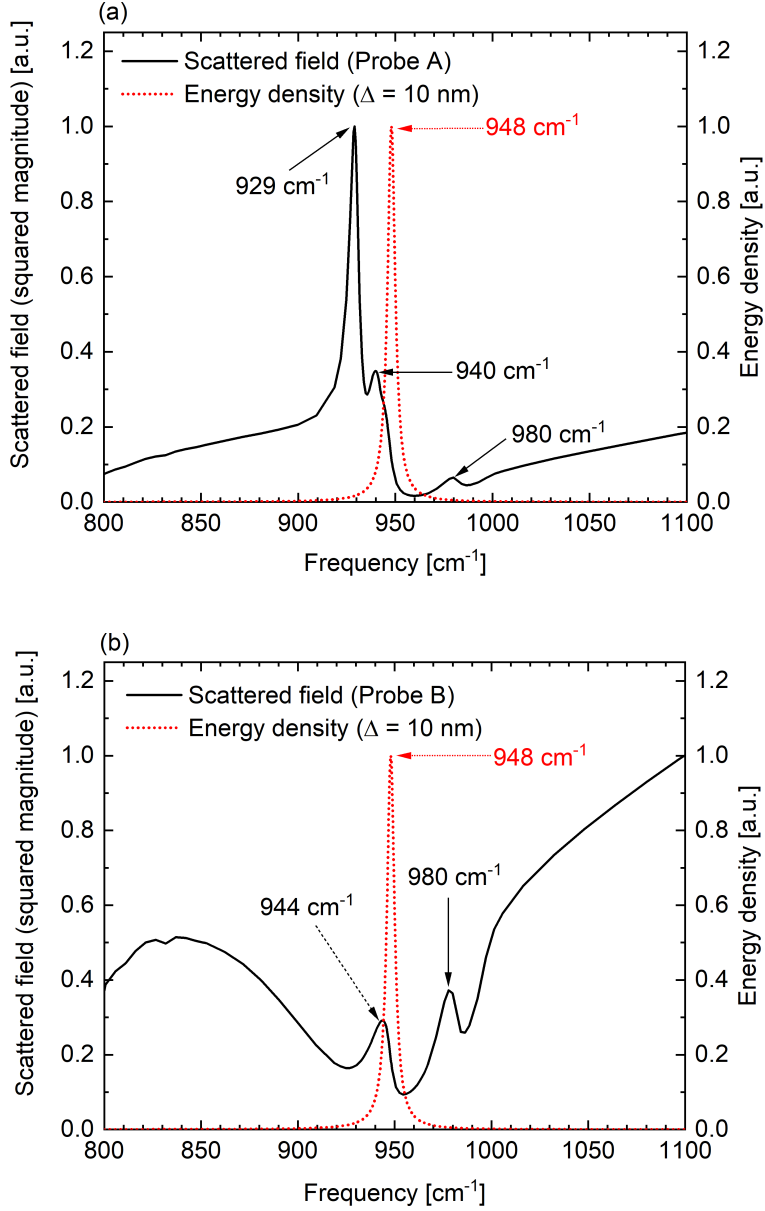

FIG. 2. Spectral distribution of scattered electric field (magnitude squared) in the far zone at an angle of 45°, with respect to the surface normal, and at a distance of 1 m: (a) Probe A, and (b) Probe B. The spectral distribution of energy density in vacuum in the absence of the probe at a distance of 10 nm with respect to the surface is plotted in both panels for comparison. All curves are normalized by their own maximum (maximum energy density: $3.98 \times 10^8$ nJ/(m$^3$cm$^{-1}$); maximum scattered field by probe A: $5.86 \times 10^{-44}$ (V/m)$^2$; maximum scattered field by probe B: $4.76 \times 10^{-45}$ (V/m)$^2$).

To demonstrate that only numerically exact fluctuational electrodynamics simulations can capture gap modes at the origin of SPhP resonance redshift, a dipole approximation is employed hereafter to calculate the field scattered by probe A. A novel version of the dipole approximation accounting for multiple reflections between the probe and the surface is derived in Section 2 of the Supplemental Material [31]. The revised expression for the scattered field (squared magnitude) in the far zone by an electric point dipole is given by:

$$
\begin{aligned}
&\left\langle \left| \mathbf{E}^{\mathrm{FF}}(\mathbf{r}) \right|^2 \right\rangle \\
&= \frac{k_0^4}{16\pi^2 r^2} \Bigg\{ \left( \cos^2\theta \left| \alpha_{p,xx}^{eff} \right|^2 \left\langle \left| E_x^{sur} \right|^2 \right\rangle + \sin^2\theta \left| \alpha_{p,zz}^{eff} \right|^2 \left\langle \left| E_z^{sur} \right|^2 \right\rangle \right) \left( 1 + \left| R^{TM} \right|^2 + 2\mathrm{Re}\left[ R^{TM} e^{2ik_0 d\cos\theta} \right] \right) \\
&\quad + \left| \alpha_{p,xx}^{eff} \right|^2 \left\langle \left| E_x^{sur} \right|^2 \right\rangle \left( 1 + \left| R^{TE} \right|^2 + 2\mathrm{Re}\left[ R^{TE} e^{2ik_0 d\cos\theta} \right] \right) \Bigg\}
\end{aligned}
\tag{7}
$$

where $\alpha_{p,\gamma\gamma}^{eff}$ is the $\gamma\gamma$-component of the effective polarizability tensor that accounts for multiple reflections within the vacuum gap separating the dipole and the surface. The effective polarizability is given by:

$$
\alpha_{p,\gamma\gamma}^{eff} = \frac{\alpha_p}{1 - \dfrac{\alpha_p \beta}{16 C_\gamma \pi \varepsilon_0 \left( d + D/2 \right)^3}}, \qquad \gamma = x, y, z
\tag{8}
$$

where $\alpha_p$ is the dipole radiative polarizability that include self-interaction (radiative reaction correction) [27], $\beta = \left( \varepsilon_s - 1 \right) / \left( \varepsilon_s + 1 \right)$, $\varepsilon_0$ is the vacuum permittivity, while $C_\gamma = 2$ for $\gamma = x, y$ and $C_z = 1$. The second term in the denominator of Eq. (8) accounts for multiple reflections between the dipole and the surface.

Figure 3(a) shows the field scattered by a 200-nm-diameter dipole, having the same size as the hemispherical tip of probe A, made of intrinsic Si. The energy density in vacuum at a distance $\Delta$ of 110 nm with respect to the surface is also plotted for comparison (corresponds to distance between dipole centroid and surface for $d$ = 10 nm). The scattered field is maximum near SPhP resonance. The slight redshift (~ 3 cm$^{-1}$) is induced by the multiple reflections between the dipole and the surface. The dipole approximation with retardation effects has been previously used to reproduce resonance redshift measured in experiments using a dipole diameter of 3.2 μm [20]. Using the same procedure, a dipole with a diameter of 6.24 μm is needed to reproduce the resonance redshift of 19 cm$^{-1}$ observed for probe A (Fig. 3(b)). The resonance redshift induced by the 6.24-μm-diameter dipole is not mediated by gap modes, which, as explained above, are at the origin of the actual redshift. The distance between the centroid of the dipole and the surface, $d + D/2$ (3.13 μm), is very large such that the second term in the denominator of Eq. (8) accounting for multiple reflections approaches zero. Therefore, electromagnetic gap modes are not generated in the gap separating the surface and the dipole. Resonance redshift and spectral broadening for the 6.24-μm-diameter dipole is explained via the spectral energy density calculated at a distance $\Delta$ of 3.13 μm with respect to the surface (see Fig. 3(b)). The energy density at 3.13 μm, unlike at 10 nm and 110 nm, is not dominated by SPhP resonance, such that the fluctuating field illuminating the dipole is broadband. This clearly shows that numerically exact simulations accounting for the actual probe geometry are needed to both capture the physics underlying the resonance redshift and to predict the amount of resonance redshift.

Note that the impacts of electromagnetic gap modes on the spectral heat rate exchanged between the surface and the probe is discussed in the Supplemental Material (see Section 3 and Fig. S1) [31].

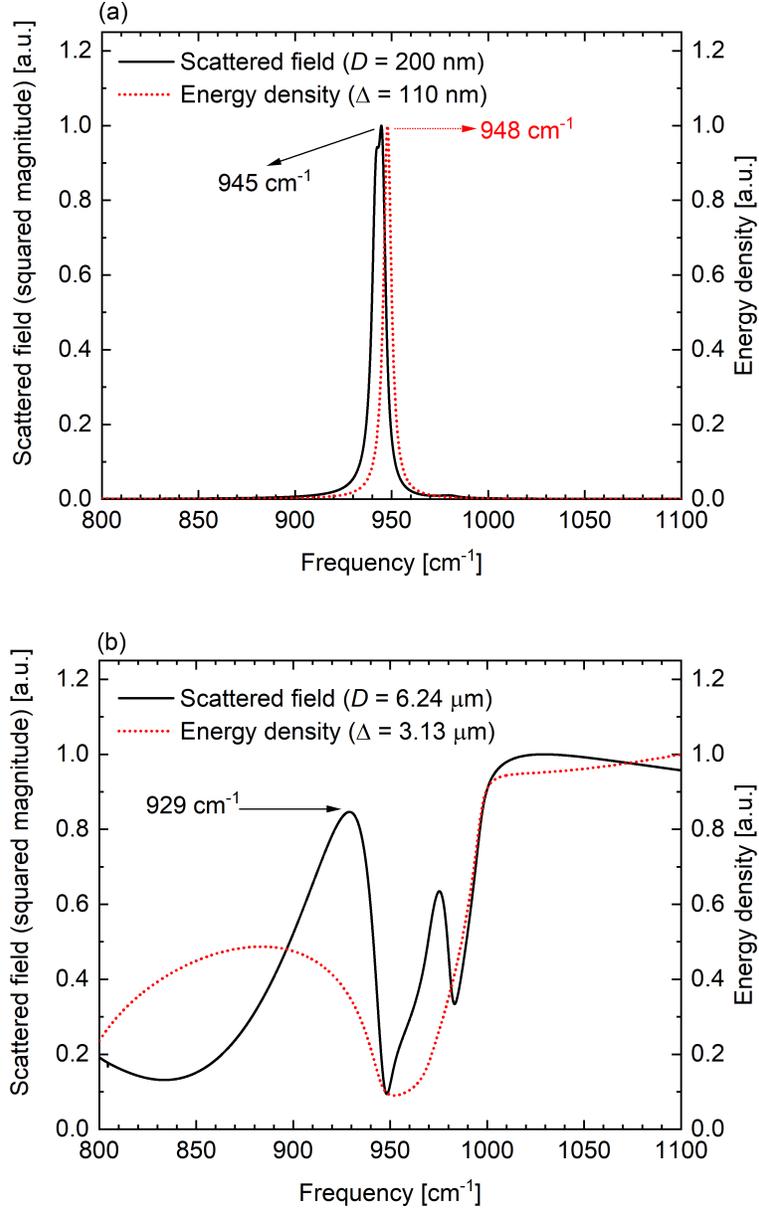

FIG. 3. Spectral distribution of scattered electric field (magnitude squared) in the far zone at an angle of 45°, with respect to the surface normal, and at a distance of 1 m: (a) 200-nm-diameter electric dipole, and (b) 6.24-$\mu$m-diameter electric dipole. The dipole edge-to-surface distance is 10 nm. Spectral distributions of energy density in vacuum at a distance of 110 nm (panel (a)) and 3.13 $\mu$m (panel (b)) with respect to the surface in the absence of a dipole are plotted for comparison. All curves are normalized by their maximum (maximum energy density and scattered field in panel (a): $3.00 \times 10^5$ nJ/(m$^3$cm$^{-1}$) and $2.49 \times 10^{-45}$ (V/m)$^2$; maximum energy density and scattered field in panel (b): $2.47 \times 10^1$ nJ/(m$^3$cm$^{-1}$) and $2.21 \times 10^{-41}$ (V/m)$^2$).

It is worth mentioning that the far-zone scattered field in an experiment is likely to exhibit additional spectral broadening when compared to numerical predictions at a fixed gap distance $d$. In actual near-field thermal spectroscopy experiments, the probing tip oscillates in a direction normal to the surface, thus resulting in a gap distance that is modulated from contact to a few tens of nanometers [20]. The spectral location of the gap modes is a function of $d/D$ (see Eq. (6)), such that modulating the gap distance generates a wide spectrum of gap modes that are further redshifted as $d/D$ decreases. Modulation of the gap distance thus results in additional spectral broadening of the far-zone scattered field due to the excitation of multiple gap modes with varying eigenfrequencies.

In summary, this work has demonstrated that resonance redshift reported in near-field thermal spectroscopy is mediated by electromagnetic gap modes induced when the probing tip size is much larger than the vacuum gap size. These gap modes affect the scattered field in the far zone even if the probe is non-resonant. Resonance redshift can be minimized with a probe with tip size approximately equal to or smaller than the vacuum gap thickness. However, this is done at the expense of measuring a scattered field in the far zone that is not dominated by resonant electromagnetic surface modes due to spectral broadening. The measured signals in near-field thermal spectroscopy experiments should be analyzed using a numerically exact framework in the probe-surface configuration, such as the T-DDA. Alternatively, numerical simulations used in tandem with the gap mode eigenfrequency expression may allow design of probe geometry and material minimizing resonance spectral redshift.

## ACKNOWLEDGMENTS

This work was sponsored by the Army Research Office under Grant No. W911NF-14-1-0210 and the University of Maine System Research Reinvestment fund. The authors acknowledge the

# REFERENCES

[1]     E. Rousseau, A. Siria, G. Jourdan, S. Volz, F. Comin, J. Chevrier, and J. J. Greffet, Nat. Photonics **3**, 514 (2009).

[2]     S. Shen, A. Narayanaswamy, and G. Chen, Nano Lett. **9**, 2909 (2009).

[3]     K. Kim, B. Song, V. Fernández-Hurtado, W. Lee, W. Jeong, L. Cui, D. Thompson, J. Feist, M. T. H. Reid, F. J. García-Vidal, J. C. Cuevas, E. Meyhofer, and P. Reddy, Nature **528**, 387 (2015).

[4]     M. P. Bernardi, D. Milovich, and M. Francoeur, Nat. Commun. **7**, 12900 (2016).

[5]     J. I. Watjen, B. Zhao, and Z. M. Zhang, Appl. Phys. Lett. **109**, 203112 (2016).

[6]     M. Ghashami, H. Geng, T. Kim, N. Iacopino, S. K. Cho, and K. Park, Phys. Rev. Lett. **120**, 175901 (2018).

[7]     C. M. Hargreaves, Phys. Lett. A **30**, 491 (1969).

[8]     D. Polder and M. Van Hove, Phys. Rev. B **4**, 3303 (1971).

[9]     R. Carminati and J.-J. Greffet, Phys. Rev. Lett. **82**, 1660 (1999).

[10]    Y. De Wilde, F. Formanek, R. Carminati, B. Gralak, P.-A. Lemoine, K. Joulain, J.-P. Mulet, Y. Chen, and J.-J. Greffet, Nature **444**, 740 (2006).

[11]    J.-P. Mulet, K. Joulain, R. Carminati, and J.-J. Greffet, Microscale Therm. Eng. **6**, 209 (2002).

[12]    C. J. Fu and Z. M. Zhang, Int. J. Heat Mass Tran. **49**, 1703 (2006).

[13]    M. Laroche, R. Carminati, and J.-J. Greffet, J. Appl. Phys. **100**, 063074 (2006).

[14]   M.P. Bernardi, O. Dupré, E. Blandre, P.-O. Chapuis, R. Vaillon, and M. Francoeur, Sci. Rep. **5**, 11626 (2015).

[15]   A. Fiorino, L. Zhu, D. Thompson, R. Mittapally, P. Reddy, and E. Meyhofer, Nat. Nanotechnol. **13**, 806 (2018).

[16]   C.R. Otey, W.T. Lau, and Fan S., Phys. Rev. Lett. **104**, 154301 (2010).

[17]   L. Tang and M. Francoeur, Opt. Express **25**, A1043 (2017).

[18]   A. Fiorino, D. Thompson, L. Zhu, R. Mittapally, S.-A. Biehs, O. Besencenet, N. El-Bondry, S. Shailendra, P. Ben-Abdallah, E. Meyhofer, and P. Reddy, ACS Nano **12**, 5774 (2018).

[19]   A. C. Jones and M. B. Raschke, Nano Lett. **12**, 1475 (2012).

[20]   A. Babuty, K. Joulain, P. O. Chapuis, J. J. Greffet, and Y. De Wilde, Phys. Rev. Lett. **110**, 146103 (2013).

[21]   B. T. O'Callahan, W. E. Lewis, A. C. Jones, and M. B. Raschke, Phys. Rev. B **89**, 245446 (2014).

[22]   B. T. O'Callahan and M. B. Raschke, APL Photonics **2**, 21301 (2017).

[23]   F. Peragut, L. Cerruti, A. Baranov, J. P. Hugonin, T. Taliercio, Y. De Wilde, and J. J. Greffet, Optica **4**, 1409 (2017).

[24]   K. Joulain, P. Ben-Abdallah, P. O. Chapuis, Y. De Wilde, A. Babuty, and C. Henkel, J. Quant. Spectrosc. Radiat. Transf. **136**, 1 (2014).

[25]   A. Jarzembski and K. Park, J. Quant. Spectrosc. Radiat. Transf. **191**, 67 (2017).

[26] S. Edalatpour, M. Cuma, T. Trueax, R. Backman, and M. Francoeur, Phys. Rev. E **91**, 063307 (2015).

[27] S. Edalatpour and M. Francoeur, Phys. Rev. B **94**, 045406 (2016).

[28] E. D. Palik, *Handbook of Optical Constants of Solids* (Academic Press, San Diego :, 1998).

[29] A. Sommerfeld, Ann. Phys. **28**, 665 (1909).

[30] A. Bãnos, Dipole Radiation in the Presence of a Conducting Half Space, Pergamon Press, Oxford (1966).

[31] See Supplemental Material at [URL will be inserted by publisher] for Section 1 and Table S1 detailing the discretization of probes A and B, for Section 2 providing a derivation of the dipole approximation with multiple reflections, and for Section 3 and Fig. S1 analyzing the heat rate between the surface and the probe.

[32] R. Schmehl, B. M. Nebeker, and E. D. Hirleman, J. Opt. Soc. Am. A **14**, 3026 (1997).

[33] K. Joulain, J. P. Mulet, F. Marquier, R. Carminati, and J. J. Greffet, Surf. Sci. Rep. **57**, 59 (2005).

[34] M. Francoeur, M. Pinar Mengüç, and R. Vaillon, J. Appl. Phys. **107**, 034313 (2010).

[35] P.K. Aravind and H. Metlu, J. Phys. Chem. **86**, 5076 (1982).

[36] P.K. Aravind and H. Metlu, Surf. Sci. **124,** 506 (1983).

[37] A. Mal'shukov, Phys. Rep. **194**, 343 (1990).

[38] V.N. Konopsky, Opt. Commun. **185**, 83 (2000).

[39]  P.I. Geshev, S. Klein, T. Witting, K. Dickmann, and M. Hietschold, *Phys. Rev. B* **70**, 075402 (2004).

[40]  A.D. Rakic, A.B. Djurisic, J.M. Elazar, and M.L. Majewski, Appl. Optics **37**, 5271 (1998).

Supplemental Material for article "Spectral redshift of the thermal near field scattered by a probe"

Sheila Edalatpour[*,†], Vahid Hatamipour[**] and Mathieu Francoeur[**,‡]

[*]Department of Mechanical Engineering, University of Maine, Orono, ME 04469, USA

[**]Radiative Energy Transfer Lab, Department of Mechanical Engineering, University of Utah, Salt Lake City, UT 84112, USA

(Dated: January 23, 2019)

[†] Corresponding author (S. Edalatpour). Tel.: + 1 207 581 2375
Email address: sheila.edalatpour@maine.edu
[‡] Corresponding author (M. Francoeur). Tel.: + 1 801 581 5721
Email address: mfrancoeur@mech.utah.edu

# 1. DISCRETIZATION OF THE PROBES

10496 and 5448 sub-volumes of varying sizes are used for discretizing probes A and B, respectively. Smaller sub-volumes are used for discretizing the portion of the probe that is closer to the surface. The number and size of the sub-volumes used for discretizing the probes are listed in Table S1.

Table S1. Size and number of sub-volumes used for discretizing probes A and B. Smaller sub-volumes are used for discretizing the portion of the probe facing the surface. The size of the sub-volumes increases as the distance with respect to the probe tip increases.

| Probe A | | | Probe B | | |
|---|---|---|---|---|---|
| Distance with respect to probe tip [nm] | Number of sub-volumes | Size of sub-volumes [nm] | Distance with respect to probe tip [nm] | Number of sub-volumes | Size of sub-volumes [nm] |
| 0 – 10 | 3804 | 2.01 | 0 – 199 | 3964 | 3.61 |
| 10 – 18 | 1172 | 4.01 | 199 – 2684 | 1484 | 57.8 |
| 18 – 30 | 756 | 5.89 | | | |
| 30 – 38 | 308 | 7.68 | | | |
| 38 – 98 | 1672 | 10.00 | | | |
| 98 – 298 | 800 | 20.00 | | | |
| 298 – 1978 | 1984 | 60.00 | | | |

## 2. DERIVATION OF DIPOLE APPROXIMATION WITH MULTIPLE REFLECTIONS

The dipole approximation is valid when the probe size is much smaller than the thermal wavelength and the separation gap. In this regime, higher order poles are not excited and uniform electric field can be assumed within the probe. The moment induced in the dipole due to thermal emission by the surface is modeled using the thermal discrete dipole approximation (T-DDA) with surface interaction as follows [1]:

$$\frac{1}{\alpha_p}\mathbf{p} - \frac{k_0^2}{\varepsilon_0}\overline{\overline{\mathbf{G}}}^R \cdot \mathbf{p} = \mathbf{E}^{sur} \tag{S1}$$

where $\alpha_p$ is the dipole radiative polarizability, $\mathbf{p}$ is the induced dipole moment, $\mathbf{E}^{sur}$ is the fluctuating field emitted by the surface, and $\overline{\overline{\mathbf{G}}}^R$ is the reflection Green's function accounting for the multiple reflections between the dipole and the surface. In the electrostatic limit, where the thermal wavelength is much larger than the dipole size and separation distance, retardation effects can be neglected [2] and the reflection Green's function reduces to the free space Green's function between the dipole and its image within the surface. The off-diagonal elements of the reflection Green's function in the electrostatic limit are equal to zero, while the diagonal elements are expressed as:

$$G_{\gamma\gamma}^R = \frac{1}{16 C_\gamma \pi k_0^2 (d+D)^3} \frac{\varepsilon_s - 1}{\varepsilon_s + 1}, \ \gamma = x, y, z \tag{S2}$$

where $d$ is the distance between the dipole edge and the surface, $D$ is the dipole diameter, $\varepsilon_s$ is the dielectric function of the surface, $C_x = C_y = 2$ while $C_z = 1$. The induced dipole moment is related to the fluctuating field by substituting Eq. (S2) into Eq. (S1):

$$\frac{1}{\alpha_{p,\gamma\gamma}^{eff}} p_\gamma = E_\gamma^{sur}, \quad \gamma = x, y, z \qquad (S3)$$

where $\alpha_{p,\gamma\gamma}^{eff}$, given by Eq. (8), is the $\gamma$-component of the effective polarizability tensor accounting for multiple reflections between the surface and the dipole. The squared magnitude of the scattered electric field, given by Eq. (7), is derived using the induced dipole moment obtained from Eq. (S3) and the far zone approximation [3].

## 3. HEAT RATE BETWEEN THE SURFACE AND THE PROBE

Near-field radiative heat transfer between a surface and a probe has been studied in the past using the T-DDA [1] and the boundary element method [4]. Yet, none of these works analyzed potential resonance redshift induced by a non-resonant probe. The heat rate between the probe and the surface is proportional to the trace of the autocorrelation function of the induced dipole moments, obtained with Eq. (3), and is given by [1]:

$$\langle Q_\omega \rangle = \frac{\omega}{2} \sum_{i=1}^{N} \left( \text{Im}\left[ \left( \alpha_i^{-1} \right)^* \right] - \frac{2}{3} k_0^3 \right) \text{tr}\left( \langle \mathbf{p}_i \otimes \mathbf{p}_i \rangle \right) \tag{S4}$$

where $\alpha_i$ is the radiative polarizability of sub-volume $i$.

Heat rate spectra for probes A and B are shown in Fig. S1. As for the scattered field, the spectral heat rate for probe A is greatly affected by gap modes and the three maxima agree reasonably well with the gap mode predictions made with Eq. (6) for $m = 0$, 1 and 2. The heat rate associated with probe B does not exhibit noticeable redshift ($\sim 1$ cm$^{-1}$). However, conversely to the scattered field, the heat rate with probe B is dominated by SPhP resonance. This is due to the fact that while the scattered field is proportional to the volume squared of the illuminated portion of the probe, the heat rate is proportional to the first power of the volume. As such, the fact that only a small portion of the volume of probe B is in the extreme near field (where SPhPs are dominant) has a smaller impact on the heat rate than the scattered field.

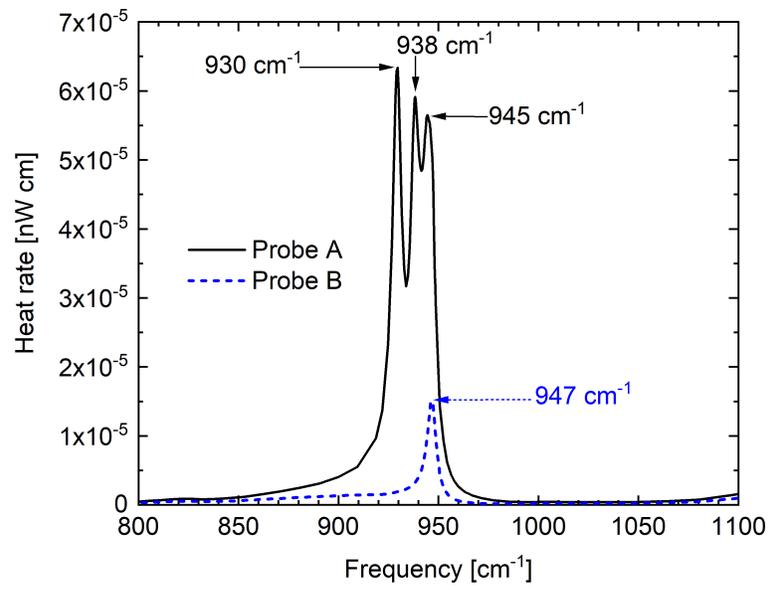

FIG. S1. Spectral distribution of heat rate with probes A and B.

# REFERENCES

[1]  S. Edalatpour and M. Francoeur, Phys. Rev. B **94**, 045406 (2016).

[2]  L. Novotny and B. Hecht, *Principles of Nano-Optics* (Cambridge University Press, New York, 2012).

[3]  R. Schmehl, B. M. Nebeker, and E. D. Hirleman, J. Opt. Soc. Am. A **14**, 3026 (1997).

[4]  K.L. Nguyen, O. Merchiers, and P.-O. Chapuis, J. Quant. Spectrosc. Radiat. Transf. **202**, 154 (2017).